\documentclass{aa}

\usepackage{graphicx}
\usepackage[normalem]{ulem}
\usepackage{txfonts}
\usepackage[modulo,switch]{lineno}
\usepackage{hyperref}  
\hypersetup{
    colorlinks=true,
    linkcolor=blue,
    filecolor=magenta,
    citecolor=blue,
    urlcolor=blue,
}
\usepackage{natbib}
\bibpunct{(}{)}{;}{a}{}{,}
\usepackage{multirow}
\usepackage{nicefrac}

\usepackage[usenames,dvipsnames]{xcolor}  %referee format not working

\definecolor{myBlue}{rgb}{.2,.7,.9}

\newcommand*\kms{\,km\,s$^{-1}$}

\begin{document}

\title{Multiple Stokes $I$ inversions to infer  
magnetic fields in the spectral range around Cr\,\textsc{i} 5782\,\AA\ }
\author{C. Kuckein\inst{1}, H. Balthasar\inst{1},  
C. Quintero Noda\inst{2,3}, A. Diercke\inst{1,4}, 
J. C. Trelles Arjona\inst{2,3}, B. Ruiz Cobo\inst{2,3},
T. Felipe\inst{2,3}, C. Denker\inst{1}, 
M. Verma\inst{1}, I. Kontogiannis\inst{1}, and M. Sobotka\inst{5} }

\authorrunning{Kuckein et al.}
\titlerunning{Multiple Stokes $I$ inversions in the vecinity of \ion{Cr}{i} 5782\,\AA}

\institute{%
    $^1$ Leibniz-Institut f{\"u}r Astrophysik Potsdam (AIP),
         An der Sternwarte 16, 
         14482 Potsdam, Germany \\
         \email{ckuckein@aip.de}\\
    $^2$ Instituto de Astrof\'{i}sica de Canarias (IAC), 
         V\'{i}a L\'{a}ctea s/n, 38205 La Laguna, Tenerife, Spain\\
    $^3$ Departamento de Astrof\'{\i}sica, Universidad de La Laguna
         38205, La Laguna, Tenerife, Spain \\ 
    $^4$ Universit\"at Potsdam,
         Institut f\"ur Physik und Astronomie,
         Karl-Liebknecht-Stra\ss{}e 24/25,
         14476 Potsdam, Germany \\
    $^5$ Astronomical Institute of the Czech Academy of Sciences (v.v.i.),
        Fri\v{c}ova 298, 25165 Ond\v{r}ejov, Czech Republic}

\date{Accepted July 22, 2021}

\abstract{% context (optional)
}
{% aims
The spectral window, containing Fraunhofer lines formed in the solar photosphere, around the magnetically sensitive \ion{Cr}{i} lines at 5780.9, 5781.1, 5781.7, 5783.0, and 5783.8\,\AA, with Land\'e $g$-factors between 1.6 and 2.5, is explored. The goal is to analyze simultaneously 15 spectral lines, which comprise \ion{Cr}{i}, \ion{Cu}{i}, \ion{Fe}{i}, \ion{Mn}{i}, and \ion{Si}{i} lines, without polarimetry to infer the thermodynamic and magnetic properties in strongly magnetized plasmas using an inversion code.}
{% methods
The study is based on a new setup at the Vacuum Tower Telescope (VTT, Tenerife) which includes fast spectroscopic scans in the wavelength range around the \ion{Cr}{i} 5781.75\,\AA\ line. The oscillator strengths $\log (gf)$ of all spectral lines, as well as their response functions to temperature, magnetic field, and Doppler velocity, are determined using the SIR code. The snapshot 385 of the Enhanced Network simulation from the Bifrost code serves to synthesize all the lines, which are in turn inverted simultaneously with SIR to establish the best inversion strategy. This strategy is then applied to VTT observations of a sunspot belonging to NOAA~12723 on 2018 September 30 and the results are compared to full-disk vector field data obtained with the Helioseismic and Magnetic Imager (HMI).}
{% results
The 15 simultaneously inverted intensity profiles (Stokes $I$) delivered accurate temperatures and Doppler velocities when compared against the simulations. The derived magnetic fields and inclinations are most accurate when the fields are oriented along the line-of-sight (LOS) and less accurate when the fields are transverse to the LOS. In general, the results appear similar to the HMI vector-field data, although some discrepancies exist. 
}
{% conclusion (optional)
The analyzed spectral range has the potential to deliver thermal, dynamic, and magnetic information in strongly magnetized features on the Sun, such as pores and sunspots, even without polarimetry. The highest sensitivity of the lines is found in the lower photosphere, on average around $\log \tau = -1$. The multiple-line inversions provide smooth results across the whole field-of-view. 
The presented spectral range and inversion strategy will be used for future VTT observing campaigns.}

  \keywords{ Sun: atmosphere  --
             Sun: photosphere  --
             Methods: data analysis  --
             Methods: numerical --
             Methods: observational --
             Techniques: spectroscopic 
             }

\maketitle

%--------------------------------------------------------------------------
\section{Introduction}
%--------------------------------------------------------------------------

Stokes profiles are sensitive to the magnetic field mainly through the Zeeman and Hanle effects. The study of polarized light using a polarimeter is the preferred and most accurate technique to infer the magnetic field vector. Some of the most successful polarimeters are currently installed at the largest solar telescopes, for example, the GREGOR Infrared Spectrograph \citep[GRIS,][]{collados12} at GREGOR \citep{schmidt12}, the CRisp Imaging Spectro-Polarimeter \citep[CRISP,][]{crisp} at the Swedish Solar Telescope \citep[SST,][]{sst},  or the Facility Infrared Spectrometer \citep[FIRS,][]{firs} at the Dunn Solar Telescope. 

Nevertheless, the use of a polarimeter brings along some limitations. Polarimetry requires a polarizing beamsplitter or a linear polarizer \citep[for example,][]{landi92,deltoroiniesta03}, reducing the light per beam on average by a factor of two. As a consequence, longer integration times are needed to achieve a good signal-to-noise ratio (SNR). Furthermore, depending on the type of polarimeter, three or four states of the retarders are needed to obtain the Stokes parameters. This leads to a slower scanning cadence by a factor of six to eight, ignoring the switching time of the retarders. This is a drawback especially for science cases where high cadences are needed to follow fast solar dynamics. Another issue comes with the sign of Stokes $Q$, $U$, and $V$, that is, the sign changes with the orientation of the magnetic field. In the quiet Sun, the spatial distribution of the magnetic field is entangled \citep[for example,][]{lites07,asensioramos14,marian2016} at spatial scales lower than the current spatial resolution of available telescopes. Therefore, in each resolution element we have contributions from many differently oriented magnetic fields. This causes self-cancellation of the Stokes $Q$, $U$, and $V$ signals, but does not influence Stokes~$I$.

The strong magnetic-field regime, defined by a Zeeman splitting which is larger than the Doppler width of the line, is reached, for example, in sunspots and pores \citep{hale1908}. The three-dimensional structure of solar active regions has been thoroughly studied over the last decades \citep[for example, ][]{lites93,westendorp01,langhans05,bellot07,beck08,balthasar08,borrero11LR,tiwari15,felipe16}. A clear picture of the general properties of sunspots and pores arose from the analysis of spectropolarimetric observations, generally interpreted with the support of detailed inversion methods that take into account the four Stokes parameters. Most spectral lines show a splitting large enough to facilitate determination of the magnetic field strength and its inclination just by analyzing Stokes $I$ profiles, that is, the intensity spectrum. The advantage of using additionally the other three Stokes parameters $Q$, $U$, and $V$ are twofold. On the one hand, it allows us to derive the polarity and the magnetic field azimuth. On the other hand, these parameters are less affected by straylight contamination due to the high intensity difference between the surrounding quiet Sun and, in general, the much lower intensity in pores and sunspots.  

Conversely, in the magnetically weak-field regime, which is usually fulfilled by spectral lines observed in quiet-Sun regions, the magnetic field can be inferred via the little broadening induced in the intensity profile by the Zeeman splitting \citep[for instance,][]{landi92,landi04}. However, in this case, line broadening has to be treated with caution since it can arise from several mechanisms: magnetic field, temperature, collision of the light-emitting particle with other particles, velocity gradients, macro- and micro-turbulence, etc \citep[for example,][]{stenflo77}. One way to separate the different contributions, and consequently to get accurate magnetic field information, is to use simultaneously a large enough number of spectral lines with high SNR \citep[for instance,][]{trelles21}. Such an optimal spectral range can be found in the visible around the \ion{Cr}{i} 5781.75\,\AA\ line, which presents a simple Zeeman triplet with large g$_\mathrm{eff}$ of 2.5 and is not blended according to \citet{harvey73}. 
This line sits in the left rising part of the curve-of-growth and exhibits a full Zeeman splitting at lower magnetic-field strengths than the commonly used iron lines at 6173\,\AA\ or 6302\,\AA.
Next to it reside other chromium lines which belong to the same multiplet but have lower g$_\mathrm{eff}$ values and more complex splitting patterns. In addition, the spectral window offers iron, manganese, silicon, and copper lines with large line strengths.  

The Vacuum Tower Telescope \citep[VTT,][]{vonderLuehe98} has a very powerful Echelle spectrograph, which provides high-spectral resolution observations. It has a new observing setup that offers simultaneous observations in three different spectral ranges, which are centered at the photospheric \ion{Cr}{i} 5781.75\,\AA\ line and the chromospheric H$\alpha$ and H$\beta$ lines at 6563 and 4861\,\AA, respectively. The setup was already successfully used in other studies \citep{verma20,kontogiannis20,abbasvand20}. However, this is the first time that the photospheric data is explored to infer information about the magnetic field using only intensity profiles. The advantage of the setup lays in the fast-scanning ability of the spectrograph, since no polarimetry is involved, which opens new possibilities to study dynamic events on the Sun with high spectral and moderate spatial resolution.

% +++++++++++++++++++++++++++++++++++++++++++++++
\section{Data} 
Two types of data are involved in this study: (1) simulations to study the behavior of the unexplored spectral lines under different magnetic configurations and (2) observations to realistically infer the physical properties of an active region on the Sun. 

% +++++++++++++++++++++++++++++++++++++++++++++++
\subsection{Numerical simulations} \label{Sect:sims}
% +++++++++++++++++++++++++++++++++++++++++++++++
We use two types of atmospheric models, 1D semi-empirical atmospheres and 3D realistic numerical simulations. In the first case, we employ the Harvard-Smithsonian Reference Atmosphere \citep[HSRA,][]{gingerich71} for computing the dependence of the intensity profile with different atmospheric parameters through the response functions (RF) \citep[for instance,][]{Landi1977} to perturbations of the atmospheric parameters. In the second case, we use snapshot 385 of the Enhanced Network simulation described in \cite{Carlsson2016} and developed with the Bifrost code \citep{Gudiksen2011}. The snapshot covers a surface of $24\times24$~Mm$^2$ with a pixel size of $48\times48$~km$^2$, while the vertical domain extends from 2.4~Mm below to 14.4~Mm above the average optical depth unity at $\lambda=5000$~\AA. The simulated scenario contains a configuration with strong network patches and quiet areas. The mentioned structures allow us to ascertain the capabilities of the spectral lines for inferring the physical information of network and internetwork features on the quiet Sun. For this work, we focus on two small areas where the magnetic field is strong and is either largely horizontal or vertical with respect to the solar surface.

\begin{figure}[!t]
 \centering
 \includegraphics[width=\hsize]{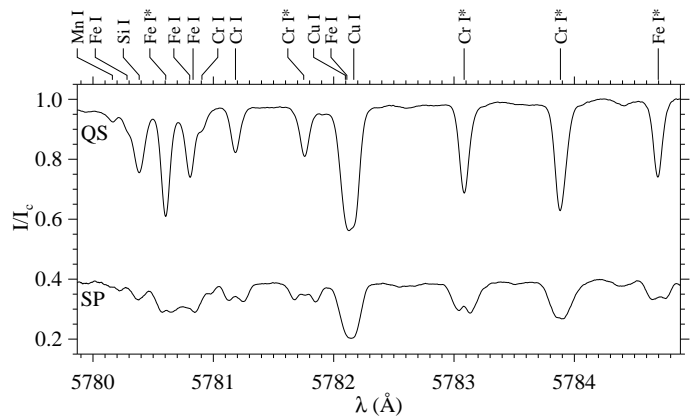}
 \caption{Observed spectra with the VTT. The upper spectrum corresponds to an average quiet-Sun profile (QS),  whereas the lower spectrum represents one profile
 within the sunspot (SP), roughly at the center of the umbra. 
 The SP spectrum is an average over five spectral pixels using a Lee-filter, available on the Interactive Data Language (IDL) with the routine \texttt{leefilt.pro}, which is an algorithm to reduce the noise. The splitting of the spectral lines is clearly visible in the strong sunspot field. Lines marked with a star (*) are used as the main lines in the inversions. All other lines are considered as blends in the SIR code. The atomic parameters for each line are listed in Table \ref{tab:atomicdata}. }
 \label{Fig:vttspectrum}
\end{figure}

\begin{table*}[!h]
\caption{Atomic data of the 15 observed spectral lines at VTT.}
\centering % used for centering table
\begin{tabular}{r c c c c c c c r} % centered columns (4 columns)

\hline\hline %inserts double horizontal lines 
Element & Wavelength (\AA) & $J_\mathrm{l}$ & $J_\mathrm{u}$ & $\log(gf)$ & $\chi_{e}$ (eV) & $g_\mathrm{eff}$ & $\alpha$ & \rule[-6pt]{0pt}{18pt} $\sigma (a_{0}^{2})$\\

\hline %inserts single line
\ion{Mn}{i} & 5780.1612 & 5.5 & 4.5 & $-1.076 \pm 0.235$ & 4.25 & $1.046$  & 0.249 & \rule{0pt}{12pt} 251 \\

%\hline %inserts single line
\ion{Fe}{i} & 5780.2828 & 4.0 & 4.0 & $-2.482 \pm 0.118$ & 4.22 & $1.325$  & 0.271 & 811 \\

%\hline %inserts single line
\ion{Si}{i} & 5780.3838 & 0.0 & 1.0 & $-2.722 \pm 0.032$ & 4.20 & $0.500$  & 0.219 &  1708\\

%\hline %inserts single line
*\ion{Fe}{i} & 5780.5994 & 3.0 & 3.0 & $-2.324 \pm 0.034$ & 3.24 & $1.625$  & 0.244 & 737\\

%\hline %inserts single line
\ion{Fe}{i} & 5780.8036 & 2.0 & 2.0 & $-2.663 \pm 0.038$ & 3.26 & $1.750$  & 0.244 &  745\\

%\hline %inserts single line
\ion{Fe}{i} & 5780.8313 & 4.0 & 5.0 & $-2.885 \pm 0.413$ & 4.43 & $2.010$  & 0.269 & 767  \\

%\hline %inserts single line
\ion{Cr}{i} & 5780.9052 & 3.0 & 2.0 & $-1.793 \pm 0.357$ & 3.32 & $1.833$  & 0.288 &  1098\\

%\hline %inserts single line
\ion{Cr}{i} & 5781.1791 & 2.0 & 1.0 & $-0.619\pm 0.029$ & 3.32 & $2.000$  & 0.288 &  1099\\

%\hline %inserts single line
*\ion{Cr}{i} & 5781.7512 & 1.0 & 0.0 & $-0.548 \pm 0.025$ & 3.32 & $2.500$  & 0.288 & 1100   \\

%\hline %inserts single line
\ion{Cu}{i} & 5782.0970 & 1.5 & 0.5 & $-1.755 \pm 0.088$ & 1.64 & $0.667$  & 0.287 &  273   \\

%\hline %inserts single line
\ion{Fe}{i} & 5782.1093 & 3.0 & 3.0 & $-0.462 \pm 0.065$ & 5.06 & $1.250$ & 0.332 &  2338 \\

%\hline %inserts single line
\ion{Cu}{i} & 5782.1670 & 1.5 & 0.5 & $-1.839 \pm 0.042$ & 1.64 & $0.667$  & 0.287 &  273   \\

%\hline %inserts single line
*\ion{Cr}{i} & 5783.0642 & 1.0 & 1.0 & $-0.184 \pm 0.025$ & 3.32 & $2.000$  & 0.288 &  1099 \\

%\hline %inserts single line
*\ion{Cr}{i} & 5783.8498 & 2.0 & 2.0 & $+0.074 \pm 0.035$ & 3.32 & $1.667$  & 0.288 &  1098 \\

%\hline %inserts single line
*\ion{Fe}{i} & 5784.6580 & 3.0 & 4.0 & $-2.369 \pm 0.031$ &  3.40 & $1.875$  & 0.244 & \rule[-2pt]{0pt}{12pt} 795    \\
\hline

\end{tabular}
%\newline   
\tablefoot{From {\it left} to {\it right}: atomic element and ionization state, wavelength, total angular momentum quantum number of the lower $J_\mathrm{l}$ and upper $J_\mathrm{u}$ levels, the oscillator strength $\log(gf)$ and its standard deviation determined in this work, the excitation potential of the lower level $\chi_{e}$, Land\'e $g$-factor, and the coefficients for collisional broadening ($\alpha$ and $\sigma)$. The abundances were extracted from \citet{Asplund2009}.
Lines marked with a star (*) are used as the main lines in the inversions, the others are considered as blends in SIR. $a_0$ is the Bohr radius. Land\'e $g$-factors were computed using LS coupling.}
 \label{tab:atomicdata}
\end{table*}

%--------------------------------------------------------------------------
\subsection{Observations and data reduction} \label{Sect:observations}
%--------------------------------------------------------------------------

The observations were acquired at the VTT with the Echelle spectrograph using a chromospheric grating with a blaze angle of 62$^\circ$ and a slit width of 80\,$\mu$m. The data was recorded simultaneously in three wavelength ranges: H$\alpha$, H$\beta$, and around the Cr\,\textsc{i} line at 5781.7\,\AA. However, here we only concentrate on the photospheric spectral range around Cr\,\textsc{i} at 5781.7\,\AA.
A pre-selection of the wavelength range and surpassing overlapping spectral orders were 
obtained with a broadband filter at $5784 \pm 5$\,\AA\ and a transmission of 74\%. 
The spectra were recorded with a pco.4000 CCD camera 
with a quantum efficiency in the range of 30\% to 40\% between 4000\,\AA\ and 6500\,\AA. 
The spectral window around the Cr\,\textsc{i} line at 5781.75\,\AA\ offers several individual photospheric lines, which are formed at similar heights in the solar atmosphere and are sensitive to the magnetic field. These lines are shown in Fig.~\ref{Fig:vttspectrum} and individually listed with their atomic data in Table \ref{tab:atomicdata}.

The observations were carried out on 2018 September 30 between 08:01~UT and 10:23~UT and included 30 raster scans of active region NOAA 12723. The active region was centered at disk coordinates (89.6\arcsec, $-$234\arcsec), which yields a $\mu \approx 0.97$, where $\mu=\cos(\theta)$ and $\theta$ is the heliocentric angle. The region comprised an emerging flux region, pores, and two small sunspots. 
The exposure time per slit position was 300\,ms. The pixel size in the scanning direction and along the slit was 0\farcs16 and 0\farcs17, respectively.
The spectral dispersion was 3.36\,m\AA\,pixel$^{-1}$ and the theoretical spectral resolution for this wavelength range is 6.98\,m\AA\,pixel$^{-1}$. 
To carry out our analysis, we selected the raster scan with the best seeing conditions, which was acquired at 10:06~UT. We focus on a region of the FOV with magnetic features of interest. Therefore, we included a small sunspot and a pore in our spatially reduced FOV of about $25\arcsec \times 20\arcsec$. 
The Kiepenheuer Adaptive Optics System \citep[KAOS,][]{kaos} was locked on small pores and operating with 26 modes. The seeing conditions were mediocre.

The data reduction involved corrections by the average dark- and flat-field images, where averages are based on individual 200 exposures. 
The average flat-field frame was also used to determine the spectrograph tilt in the images. This was accomplished
by examining the tilt in the spectral lines and was corrected by shearing the frames until the tilt
disappeared.

To perform the wavelength calibration of the spectrum, we used the cores of three lines in the averaged flat-field spectra: the \ion{Fe}{i} line at 5780.59\,\AA, the central \ion{Cr}{I} line at 5781.75\,\AA, and the \ion{Cr}{i} line at 5783.84\,\AA. We fitted a second order polynomial to the spectral lines to determine the exact line-core positions. The relative difference of the line-core position is compared to the actual wavelength positions in the Fourier transform spectra \citep[FTS;][]{neckel84}. After a first approximation of the wavelength dispersion, the FTS spectrum and the VTT spectrum are correlated to calculate the exact wavelength dispersion of the spectrum.

The continuum of the spectra was normalized carrying out the following steps: (1) we computed a mean quiet-Sun (QS) profile by averaging a $120 \times 230$ pixels area of granulation in the original, much larger, FOV of $660 \times 442$ pixels; (2) from the mean QS profile we averaged 30 spectral points corresponding to the continuum around 5784.2\,\AA, located between the rightmost \ion{Cr}{i} and \ion{Fe}{i} lines; (3) all spectra were then divided by this continuum value; (4) the resulting continuum shows a slight dependence with wavelength, likely owing to instrumental effects. This was solved by comparing the average quiet-Sun profile at several continuum wavelength positions with the FTS continuum at the same wavelength positions. As a result, we introduced a wavelength-dependent normalization factor to achieve that the continuum of the average QS profile matches the FTS's one. 
The calculated QS profile served as well for the determination of the oscillator strengths (see Sect.~\ref{Sect:loggf}) and as a straylight profile for the inversion code (see Sect.~\ref{Sect:VTTinversions}). 

\begin{figure*}[!t]
 \centering
 \includegraphics[width=\hsize]{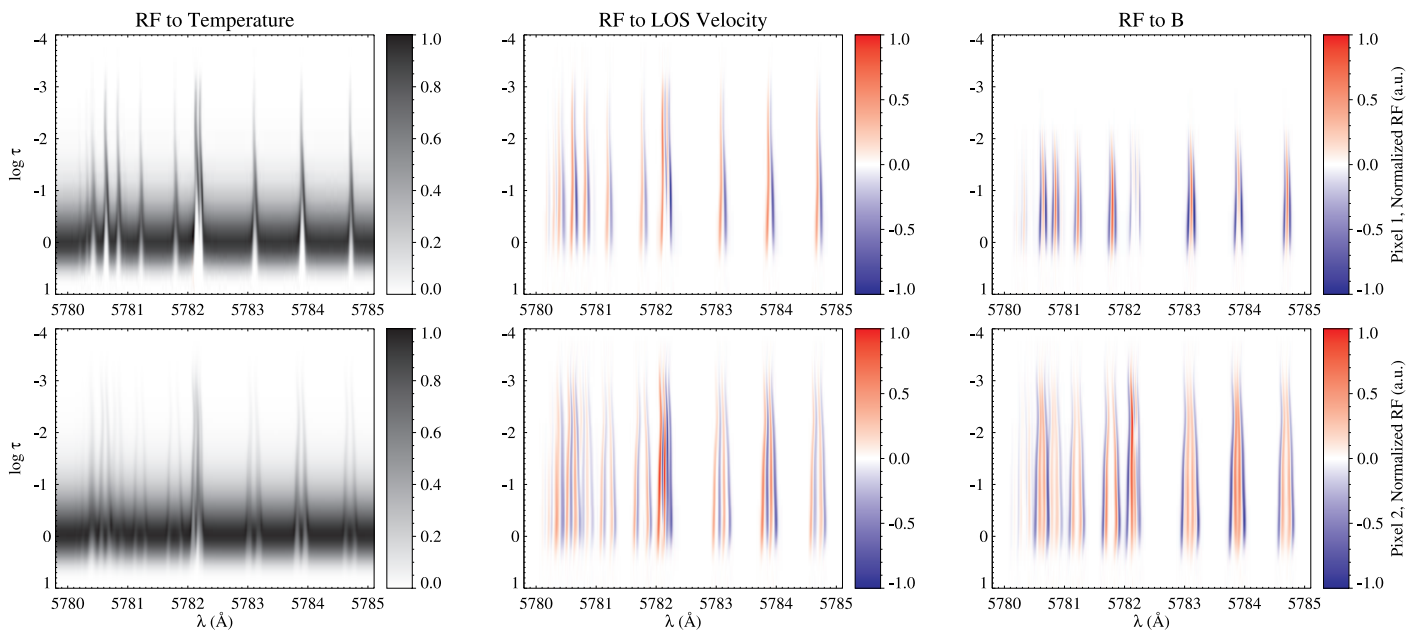}
 \caption{Response functions of Stokes $I$ to changes of the temperature, LOS velocity, and magnetic field strength for all spectral lines of Table \ref{tab:atomicdata}. Pixel~1 (\emph{top row}) corresponds to a position of the simulation cube with a more inclined magnetic field of $B=535$\,G and $\gamma = 132^\circ$ at $\log \tau = 0$. Pixel 2 (\emph{bottom row}) represents a more vertical and strong magnetic field of $B=2800$\,G and $\gamma = 164^\circ$ at $\log \tau = 0$.}
 \label{Fig:RF}
\end{figure*}

Magnetograms and continuum images from the Helioseismic and Magnetic Imager \citep[HMI,][]{hmi} on board of the Solar Dynamics Observatory \citep[SDO,][]{sdo} were used to align and compare the magnetic field with the one inferred from the VTT data. HMI acquires filtergrams in six positions (times four Stokes parameters) of the photospheric \ion{Fe}{i} 6173\,\AA\ line. We used the HMI full-disk vector field data (hmi.ME\_720s\_fd10), which are 12-minute averages, to extract the information of the flux and line-of-sight (LOS) inclination of the magnetic field \citep{hoeksema14}. The data were then rotated and scaled to match the FOV of the VTT observations. 

%--------------------------------------------------------------------------
\section{Method and analysis} \label{Sect:methods}
%--------------------------------------------------------------------------

The magnetic field causes a splitting in most spectral lines (Zeeman-effect). The width of this splitting depends on the strength of the magnetic field. The $\pi$-components are linearly polarized and are strongest when looking perpendicular to the magnetic field. Looking along the field lines, $\pi$-components disappear. Thus, the Stokes $I$-profile contains information about the magnetic field  strength $B$ and its inclination $\gamma$, but sign-independent. Fitting the whole $I$ profile of one or more spectral lines with the SIR code \citep[Stokes inversion based on Response Functions;][]{SIR} allows to infer $|B|$ and $\gamma$. A comparison with, for example, HMI-data resolves the polarity ambiguity for $\gamma$. Using several lines simultaneously significantly helps to improve the fits and reduce the ambiguities. This will be discussed below. 
A similar method, but with \ion{Fe}{i} lines above 6000\,\AA, was introduced by \citet{BS93}.
However, the authors treated the lines independently and did not infer 
information about inclination of the magnetic field with respect to the LOS.

The radiative transfer equation establishes that intensity profiles depend on the magnetic field only through the absorption matrix. Neglecting the dependence on Stokes $Q$ and $U$, because in general Stokes $I$ is much larger than $Q$ and $U$, we can assume that the intensity perturbation of the outcoming radiation is given by $\delta I \sim \eta_I I + \eta_V V$. Equation~\ref{Eq:absorption} \citep[see for instance][]{landi76}  shows the dependence of the absorption coefficient $\eta_V$ on the cosine of the magnetic field inclination $\gamma$, and $\eta_I$ on the square of the sine of $\gamma$, and both coefficients on $\eta_p$, $\eta_b$, and $\eta_r$, the Voigt profiles weighted with the Zeeman amplitudes for linear, right, and left circular polarization, respectively: 
\begin{align} 
   \label{Eq:absorption}
   \eta_I &= 1/2[\eta_p \sin^2(\gamma) + 1/2 (\eta_r+\eta_b) (2-\sin^2(\gamma)] \\
   \eta_V &= 1/2(\eta_r-\eta_b) \cos(\gamma) \nonumber
\end{align}
If we change the magnetic field polarity, both $V$ and $\eta_V$ 
change sign, producing the same effect on Stokes $I$.  Consequently, analyzing only Stokes~$I$ allows us the unambiguous determination of the magnetic field strength and of $\sin(\gamma)$. However, Stokes $I$ is sensitive neither to the azimuth nor to the polarity of the field.

% +++++++++++++++++++++++++++++++++++++++++++++++
\subsection{Determination of $\log$\,$(gf)$} \label{Sect:loggf}
% +++++++++++++++++++++++++++++++++++++++++++++++

For a correct characterization of a spectral line, it is crucial to know its oscillator strength $f$ \citep{trelles21}. This value is defined as the ratio between the probability of absorption or emission of electromagnetic radiation in a transition between energy levels of an atom or molecule and the probability of this transition in a classical oscillator \citep{mihalas78}. Traditionally it is expressed together with the statistical weight $g$ of the level. 

First attempts using the few $\log$\,$(gf)$ values found in the literature to invert the observations did not provide good spectral-line fits. Even with standard atmospheric models, trying to fit individual spectral lines caused that the fit to the remaining lines got worse. This is evidence that the $\log$\,$(gf)$ values needed to be revised to enhance their accuracy. Therefore, we decided to compute them on our own. It is possible to carry out this task by comparing the synthetic and observational spectra \citep{gurtovenko81,gurtovenko82,thevenin89,thevenin90,borrero03}. This was done for all spectral lines. The derived values are listed in Table \ref{tab:atomicdata}. For those lines with previous-existing values obtained from the literature, the new $\log$\,$(gf)$'s are not far away from the original values, that is, they are found within some margin of uncertainty. Nevertheless, these small changes sufficiently enhance the quality of the fitted lines when fitting them all together during the inversions. 

The same steps as described by \citet{trelles21} were followed to infer the $\log(gf)$ parameters. This method consists of an iterative process to retrieve the best combination of $\log(gf)$ values and model atmospheres to match the synthesized and observed spectral profiles. The abundances were extracted from 
\citet{Asplund2009}.
In the first step, the $\log(gf)$ values were computed for the following five spectral lines: \ion{Cr}{i} at 5781.18, 5781.75, 5783.06, 5783.85\,\AA, and \ion{Fe}{i} at 5784.66\,\AA. In that process, we used the average quiet-Sun spectra from the VTT observations and modeled the spectral lines with an initial FAL-C model atmosphere \citep{fontenla93}. In the second step, the previously determined $\log(gf)$ values were used to invert the data for all pixels of the observed region. The obtained model atmospheres were then used to compute the $\log(gf)$ values of the remaining spectral lines. In the third step, we repeated the pixel by pixel inversions using the derived $\log(gf)$ values of all spectral lines to re-determine them.  Finally, $\log(gf)$ and model atmospheres were inferred pixel by pixel. This process allows us to calculate the mean and standard deviation of $\log(gf)$ for each spectral line. The results are listed in Table~\ref{tab:atomicdata}. Note that the values for the Cu-lines are not pure $\log(gf)$-values. The given values reflect also the relative abundances of the isotopes (see Sect.~\ref{Sect:Cu_line}).

% +++++++++++++++++++++++++++++++++++++++++++++++
\subsection{Spectral synthesis} \label{Sect:spectral_synth}
% +++++++++++++++++++++++++++++++++++++++++++++++

We make use of the SIR code for synthesizing the full Stokes vector of the spectral lines displayed in Table \ref{tab:atomicdata}, which arise from the numerical simulations described in Sect. \ref{Sect:sims}. 
Disk-center observations ($\mu=1$) are assumed. The abundance values of the different atomic species are given in \cite{Asplund2009} and are 7.51, 5.64, 5.43, 7.50, and 4.19 dex for Si, Cr, Mn, Fe, and Cu, respectively. 
We define a spectral sampling of 3.36~m\AA \ matching that of the VTT observations, which we will analyze later. 
We do not include microturbulence enhancement in the synthesis. 
Spectral degradation based on the VTT results were taken into account. These effects include straylight (15\,\%), instrumental line broadening (a Gaussian with a FWHM of $\sim$1.1\kms) and noise comparable to the noise found in the quiet-Sun continuum of the VTT spectra.

% +++++++++++++++++++++++++++++++++++++++++++++++
\subsection{Inversions} \label{Sect:inversions}
% +++++++++++++++++++++++++++++++++++++++++++++++

The spectral-line inversions were also carried out with SIR. The code assumes local thermodynamical equilibrium (LTE) and hydrostatic equilibrium to iteratively solve the radiative transfer equation. The atomic data appear in Table \ref{tab:atomicdata} and were taken from either the National Institute of Standards and Technology \citep[NIST;][]{NIST} or the Vienna Atomic Line Data Base \citep[VALD,][]{piskunov95,vald}. The solar abundances were retrieved from \citet{Asplund2009} and the broadening of the spectral lines by collisions with neutral hydrogen atoms was computed using the Anstee, Barklem and O'Mara (ABO) theory \citep[for example,][]{Anstee1995, Barklem1997}. To this end, the ABO-cross calculator code \citep{Barklem2015} was used, which interpolates in pre-computed tables of line broadening parameters.

The inversion code provides height-dependent physical parameters.  
The height is expressed in units of the logarithm of the LOS continuum
optical depth at 5000\,\AA. 
The inversion strategy was as follows: (1) We divided the entire spectral range into five spectral blocks. Each block has one main spectral line (marked with a star symbol (*) in Fig. \ref{Fig:vttspectrum} and Table \ref{tab:atomicdata}) and the others were considered as blends in the SIR code. Hence, we inverted simultaneously 15 intensity profiles. 
(2) The following physical variables were set as free parameters in the inversions: temperature $T$, microturbulence $v_\mathrm{mic}$, magnetic field strength $B$, LOS inclination $\gamma$, and LOS velocity $v$. The relevant optical depths covered by all the lines is studied in the next section.  

We perform the inversions using the upgraded version of SIR described in Section 5 of \cite{Gafeira2021}. This upgrade is available in a public online repository\footnote{\url{https://gitlab.com/gafeira/parallel_desire_sir_rh}} and facilitates running the code in parallel distributing the inversion of pixels between the CPU cores available on the computer.

% +++++++++++++++++++++++++++++++++++++++++++++++
\subsection{Response functions} \label{Sect:RFs}
% +++++++++++++++++++++++++++++++++++++++++++++++

The optical depth, to which the spectral lines are sensitive to for different physical quantities, is computed using response functions \citep[RFs,][]{ruizcobo94}. We used two different pixels from the simulation cube: pixel 1 corresponds to moderate and more inclined magnetic fields, whereas pixel 2 shows strong, rather vertical magnetic fields. The pixels are representative of phenomena on the Sun with strong fields, like pores or sunspots. Figure \ref{Fig:RF} exhibits the response functions of Stokes $I$ to temperature, LOS velocity, and magnetic field changes, for all the lines listed in Table \ref{tab:atomicdata}. As expected from the $g_\mathrm{eff}$ values, all involved lines are sensitive to the magnetic field, especially between $\log \tau \sim$ 0 and $-1.5$. 
The response functions also reveal that the more inclined the magnetic fields are, the lower is the sensitivity at smaller optical depths (upper right panel). All spectral lines are sensitive to changes in the temperature and LOS velocity, although differences exist among the two analyzed pixels. For instance, in pixel 2, which has more vertical magnetic fields, the spectral lines become less sensitive to temperature changes with $\log \tau \lesssim$\,$-1.5$ (lower left panel). 

We conclude that all 15 spectral lines roughly cover the same optical depths, which gives us confidence that the lines will complement each other during the inversion process. Hence, the lines will minimize ambiguities which can arise from small changes in, for example, the magnetic field or the LOS velocity.

% +++++++++++++++++++++++++++++++++++++++++++++++
\subsection{Spectral lines of Cu isotopes} \label{Sect:Cu_line}
% +++++++++++++++++++++++++++++++++++++++++++++++
The \ion{Cu}{i} lines in Table \ref{tab:atomicdata} need special attention among all the involved lines. They appear as a single line associated to a wavelength of 5782.132\,\AA\ in the NIST catalogue. However, the VALD data base lists two lines with equal atomic data. Hence, the lines could arise from different isotopes. There are two stable isotopes: $^{63}$\ion{Cu}{} and $^{65}$\ion{Cu}{} with relative abundances of 69.17\% and 30.83\%, respectively.
In the following paragraph, we investigated whether one or two \ion{Cu}{i} lines should be considered during the inversions.

\begin{figure}[!t]
 \centering
 \includegraphics[width=\hsize]{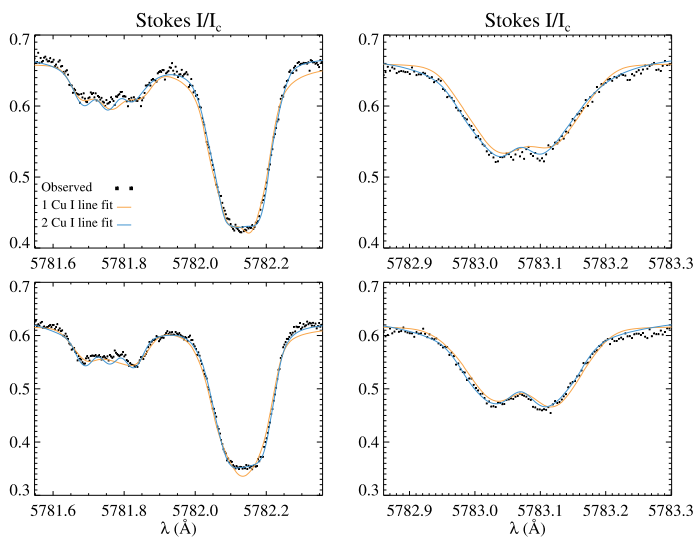}
 \caption{Inversion results taking into account one or two \ion{Cu}{i} lines. The upper and lower rows show the spectra of two different pixels within the sunspot data of the VTT. The dots represent the observations, whereas the orange and blue solid lines show the best fit from the inversion for one or both \ion{Cu}{i} lines, respectively. The left panels show the \ion{Cr}{i} 5781.75\,\AA\ Zeeman-splitted line and next to it on the right the \ion{Cu}{i} line(s). The right panel shows the impact of having one or two \ion{Cu}{i} lines on the \ion{Cr}{i} 5783.06\,\AA\ line.}
 \label{Fig:CuI}
\end{figure}

First, the wavelength for each \ion{Cu}{i} line was determined by equidistantly separating them a fixed amount from the central wavelength (5782.132\,\AA). This was done systematically by synthesizing the line with SIR and comparing it with the average quiet-Sun profile from the observations. The best result was obtained when shifting the lines $\Delta \lambda$$=$$\pm$0.035\,\AA. The resulting wavelengths are listed in Table \ref{tab:atomicdata}. Second, we performed test inversions with SIR to measure the impact on the fits when using one or two \ion{Cu}{i} lines. The results are displayed in Fig. \ref{Fig:CuI}. The upper and lower rows show the spectra of two selected pixels inside of the sunspot, from the VTT observations. In the left panels we see the prominent Zeeman splitting of the \ion{Cr}{i} 5781.75\,\AA\ line and on the righthand side the \ion{Cu}{i} lines blended also by the \ion{Fe}{i} 5782.10\,\AA\ line. The observations are represented by black dots. The orange solid line shows the best fit from the multi-line inversions, including all lines of Table \ref{tab:atomicdata}, but only one \ion{Cu}{i} line. The results including all lines, that is, also both \ion{Cu}{i} lines, is depicted with the blue solid line. To also see the impact on more distant spectral lines we included the right panel in Fig. \ref{Fig:CuI}, which shows another \ion{Cr}{i} line more to the red, at 5783.06\,\AA. The results with two \ion{Cu}{i} lines (blue line) fit better the observed spectra. Especially the line core of \ion{Cu}{i} in the lower-left panel is better fitted with two \ion{Cu}{i} lines. In addition, the wings of all lines match better in the case of using two \ion{Cu}{i} lines. The same applies for the \ion{Cr}{i} line in the righthand panel. 

We conclude that treating the \ion{Cu}{i} lines as separate transitions improves the quality of the inversions. Hence, we will use two \ion{Cu}{i} lines for the analysis described throughout the manuscript.    

%--------------------------------------------------------------------------
\section{Results}
%--------------------------------------------------------------------------

The main goal is to proof the reliability of inferring the magnetic field strength in the present spectral range using simultaneously many different intensity profiles. Our rationale is founded on the expected improvement of the inferred magnetic field thanks to the additional information provided by a high number of spectral lines. In the process of inverting the spectral lines, we will also infer other physical parameters such as the temperature, the LOS velocity, and the magnetic field inclination. In the following subsection, we will first test our method on synthetic intensity profiles from the simulation cube described in Sect. \ref{Sect:sims}. This is an essential step to verify that the inferred results with an inversion code match the known ones of the simulation cube. In the second part, we will then apply what we learned from the simulations to invert the observations from the VTT described in Sect. \ref{Sect:observations}.

%--------------------------------------------------------------------------
\subsection{Inversion of synthetic spectra from the simulations} \label{Sect:InvSimulations}
%--------------------------------------------------------------------------

\begin{figure}[!t]
 \centering
\includegraphics[width=\hsize]{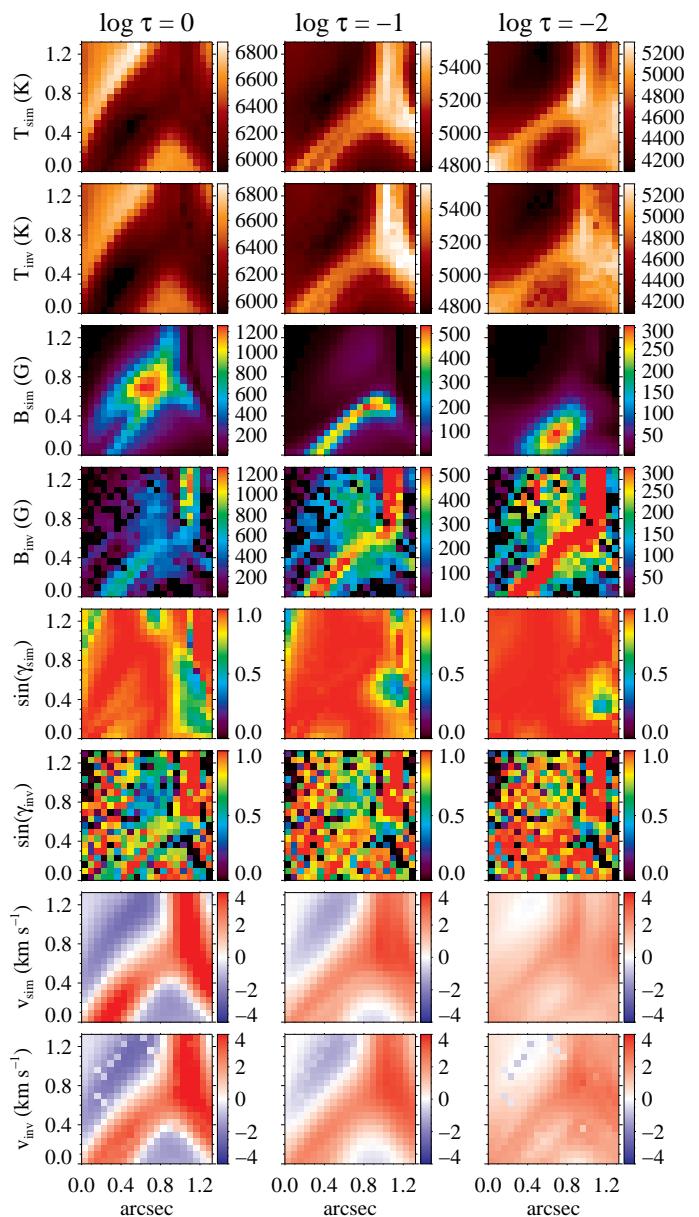}
 \caption{Comparison of physical parameters between the simulations and the SIR inversions. The cutout area A represents a region of strong horizontal magnetic fields. From top to bottom: temperature $T$, magnetic field strength $B$, sine of the inclination $\sin (\gamma)$, and LOS velocity $v$. The first row of each physical parameter corresponds to the simulation cube, whereas the second row is the result from the inversions. From left to right: three different optical depths ascending in height: $\log \tau =$\,0, $-1$, and $-2$. Note that the color scales are variable among the different optical depths in the upper four rows. }
 \label{Fig:sim_inv_area_a}
\end{figure}

\begin{figure}[!t]
 \centering
 \includegraphics[width=\hsize]{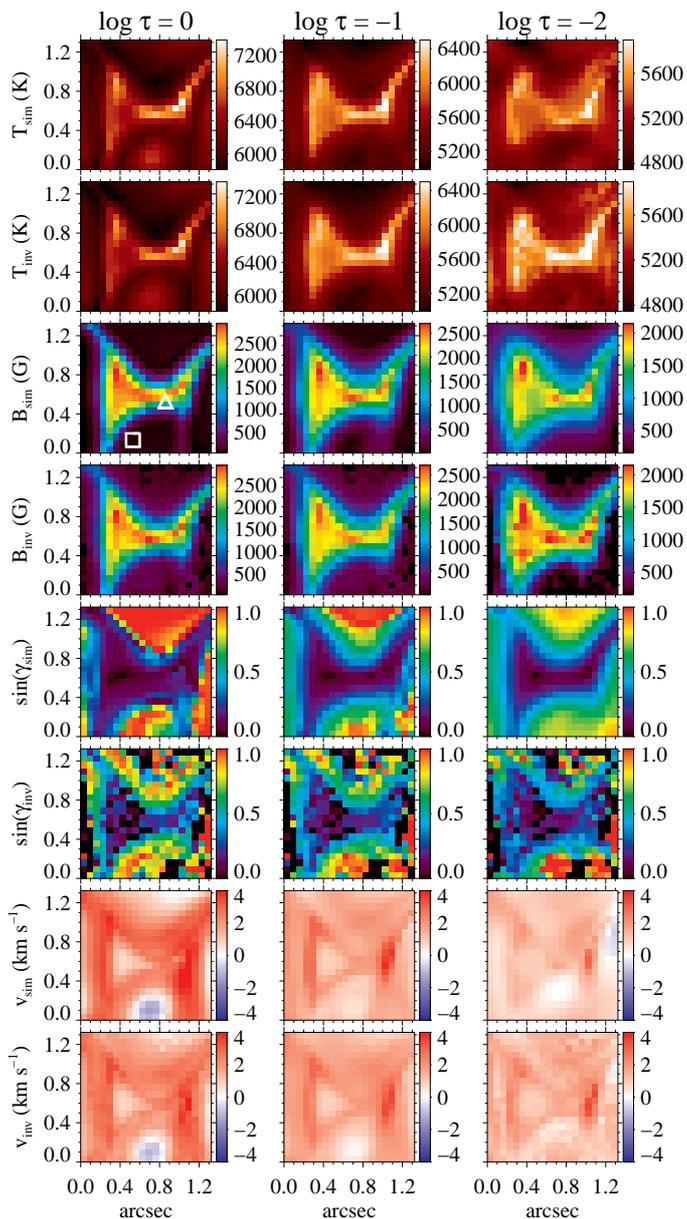}
 \caption{Same as Fig. \ref{Fig:sim_inv_area_a} but for area B, which comprises strong vertical magnetic fields. The atmospheric stratification of two pixels marked with a square and triangle in the lefthand magnetic field panel is shown in Fig.~\ref{Fig:atmos_tau}.}
 \label{Fig:sim_inv_area_b}
\end{figure}

We extracted two $1\farcs3 \times 1\farcs3$ ($20 \times 20$ pixel) areas from the large simulation cube presented in Sect. \ref{Sect:sims}. The two areas A and B were selected because they harbor the strongest horizontal and vertical magnetic fields within the cube. Therefore, they sample strong magnetic-field structures with prominent Zeeman-splitting patterns, comparable to pores and sunspots on the Sun. Figures \ref{Fig:sim_inv_area_a} and \ref{Fig:sim_inv_area_b} show an overview of the temperature $T$, magnetic field strength $B$, sine of the magnetic field inclination $\sin (\gamma)$, and LOS velocity $v$ of areas A and B, respectively.
As explained in Sect. \ref{Sect:methods}, Figures \ref{Fig:sim_inv_area_a} and \ref{Fig:sim_inv_area_b} only show $\sin (\gamma)$ instead of the inclination, because we cannot derive the polarity of the field. The physical parameters are shown for three selected optical depths: $\log \tau =$\,0, $-1$, and $-2$.

We carried out parallel SIR test inversions with a large set of different configurations to find the optimal initial parameters. These tests included variations in the amount of cycles, that is, inversion runs taking into account the result of the previous inversion), nodes along the atmosphere, and number of initial atmospheres. The goal is to keep the amount of nodes as low as possible while inferring atmospheres which resemble closely the ones from the simulations. The synthetic intensity profiles were degraded owing to straylight, instrumental line broadening, and noise, as already described in Sect.~\ref{Sect:spectral_synth}. 

An optimal set of parameters for SIR, when inverting simultaneously our 15 spectral lines, was found for: two cycles, where the first cycle had a fixed amount of lower nodes and the second cycle used the automatic selection of nodes. A summary of the nodes appears in Table~\ref{tab:nodes}. The azimuth $\phi$ was neglected since no information can be retrieved from only intensity profiles. 

\begin{table}[!b]
\begin{center}
\caption{Inversion strategy for the multi-line intensity inversions of the simulations. The nodes for the parameters with a (*) were automatically selected by SIR in the second cycle. Therefore, an upper limit of the node is indicated. }\label{tab:nodes}
\begin{tabular}{l|cr}
\hline
\hline
\multirow{2}{*}{Parameter}     & \multicolumn{2}{c}{Nodes}  \rule[-4pt]{0pt}{15pt}  \\
                & Cycle 1      &  Cycle 2                            \\
\hline
\hline
Temperature*       & 3  &  $\leq 5$ \rule[-4pt]{0pt}{14pt} \\
Microturbulence    & 1  &  1         \\
Magnetic field*    & 1  &  $\leq 2$ \\
Inclination*       & 1  &  $\leq 2$ \\
Azimuth            & 0  &  0         \\
Velocity*          & 1  &  $\leq 2$ \\
\hline
\end{tabular}
\end{center}
\end{table}

We noticed a better performance when using many initial-guess atmospheres. This has been already used in several other studies \citep[for example,][]{quinteronoda14,kuckein19}. The closer the initial atmosphere is to the ``real" atmosphere on the Sun, the quicker and better are the results of the inversion code.  
We found good results using eight initial atmospheres, which comprised 73 optical-depth positions between $2.4$\,$\geq$\,$\log \tau$\,$\geq$\,$-4.8$. The atmospheres were based on the FAL-C atmospheric model, where we introduced linear variations of the temperature, LOS velocity, and the magnetic field vector. 
We used the $\chi^2$ value, that is, the sum of the squared difference between the degraded-simulated and the synthesized intensity profiles, to keep the best inversion out of the eight attempts for each pixel.

The results are depicted in Figs. \ref{Fig:sim_inv_area_a} and \ref{Fig:sim_inv_area_b}, for the case of strong horizontal (area A) and strong vertical (area B) magnetic fields, respectively. For area A we find a very good match between the simulations and the inversions for the temperature and LOS velocities, across all the shown optical depths. The magnetic field is not well matched. The inversions significantly underestimate the field strength in the central pixels for $\log \tau = 0$, about 300--400\,G less. At $\log \tau = -1$ the match in field strength between simulations and inversions is better, although many pixels, especially in the upper right part, show false enhanced fields which are not seen in the simulations. There is a large mismatch of the magnetic fields for $\log \tau = -2$. This is not surprising, since the response function to more inclined, that is, horizontal fields, vanishes for this optical depth (upper right panel in Fig. \ref{Fig:RF}). The results for the inclination show that the inversion code often fails to retrieve reliably the inclination of the pixels. 

\begin{figure}[!t]
 \centering
 \includegraphics[width=\hsize]{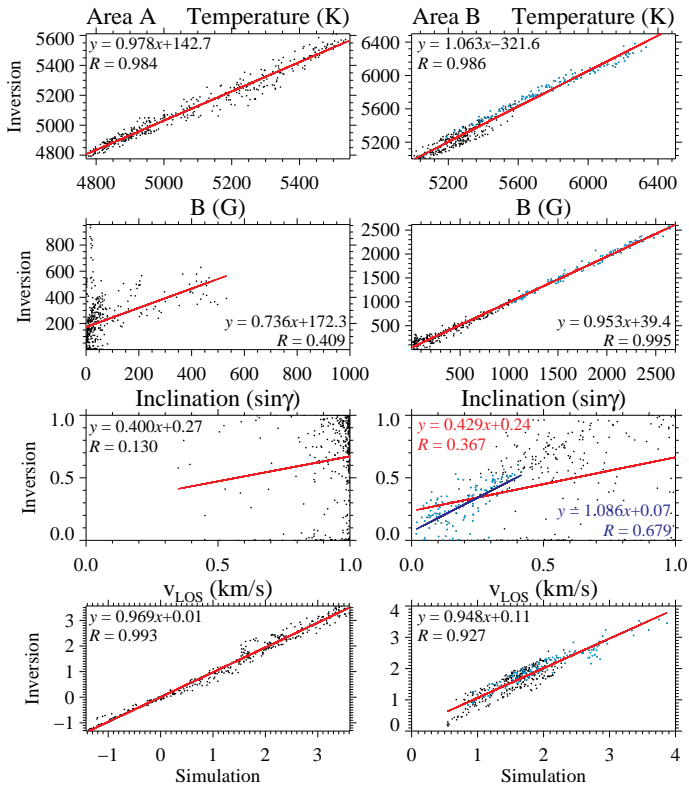}
 \caption{Scatter plots showing the relationship between physical parameters in the simulations versus in the inversions of areas A (\emph{left}, strong horizontal fields) and B (\emph{right}, strong vertical fields) for $\log \tau = -1$. The light-blue dots in area B correspond to pixels with $B \geq 1000$\,G. The correlation coefficient is given by the $R$ value. The equation $y$ represents the red solid line, which is a linear fit to all the points. In the inclination panel of area B the blue solid line fits only the light-blue points and the associated equation $y$ appears in blue. }
 \label{Fig:scatterplots}
\end{figure}

\begin{figure}[!h]
 \centering
 \includegraphics[width=\hsize]{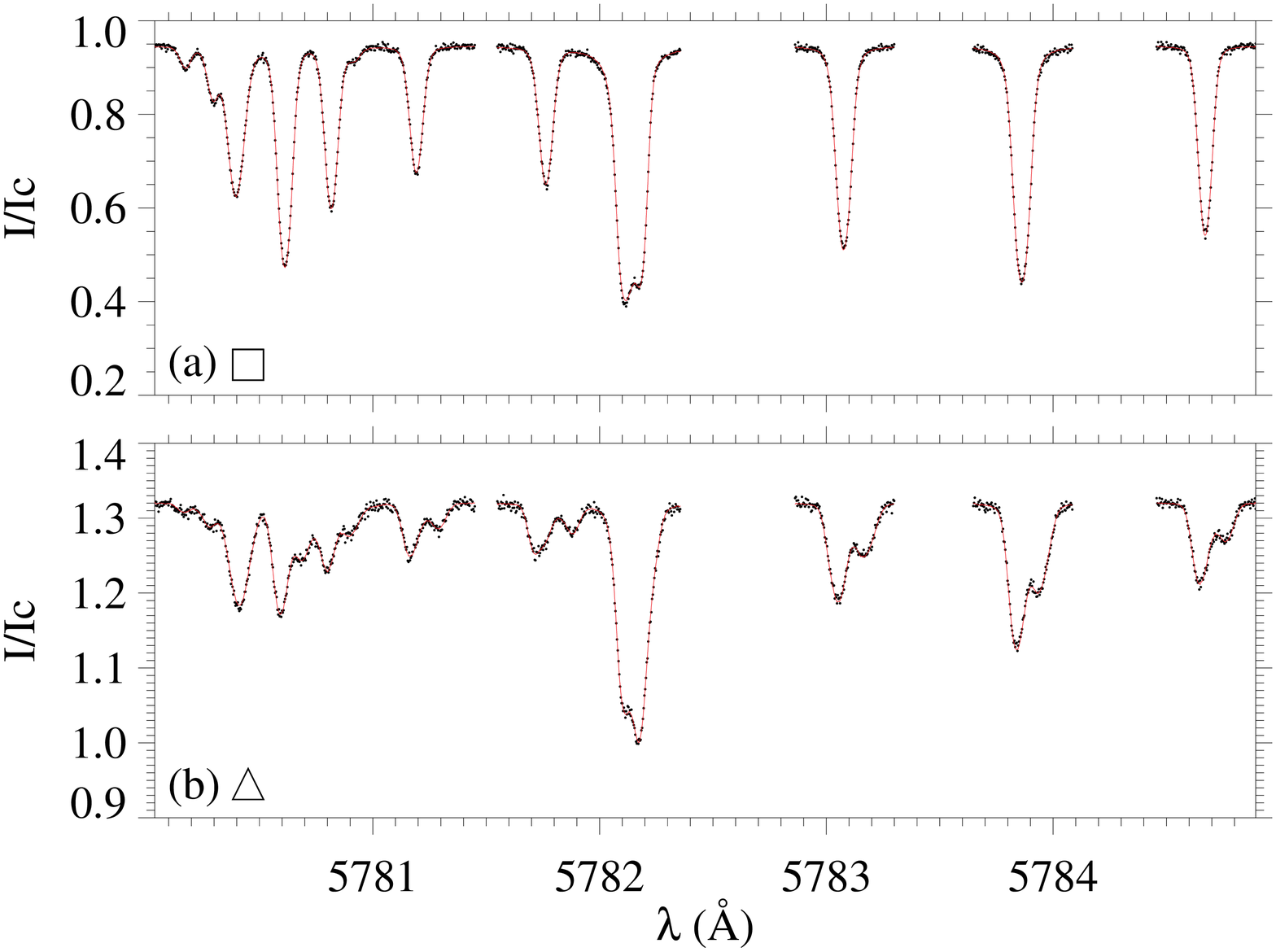}
 \caption{Synthesized spectra (black dots) from the simulations and result of the inversions (solid red line). Pixel (a) ({\it top} panel) corresponds to an example of a low magnetic field, which is located at ($x,y$) = (0.53\arcsec, 0.13\arcsec) in Fig. \ref{Fig:sim_inv_area_b} (square symbol). Pixel (b) (triangle symbol, {\it bottom} panel) is located at ($x,y$) = (0.86\arcsec, 0.53\arcsec) and represents a pixel with a strong magnetic field. The associated atmospheric stratification is shown in Fig.~\ref{Fig:atmos_tau}.}
 \label{Fig:profiles}
\end{figure}

\begin{figure*}[!h]
\sidecaption
  \includegraphics[width=12.5cm]{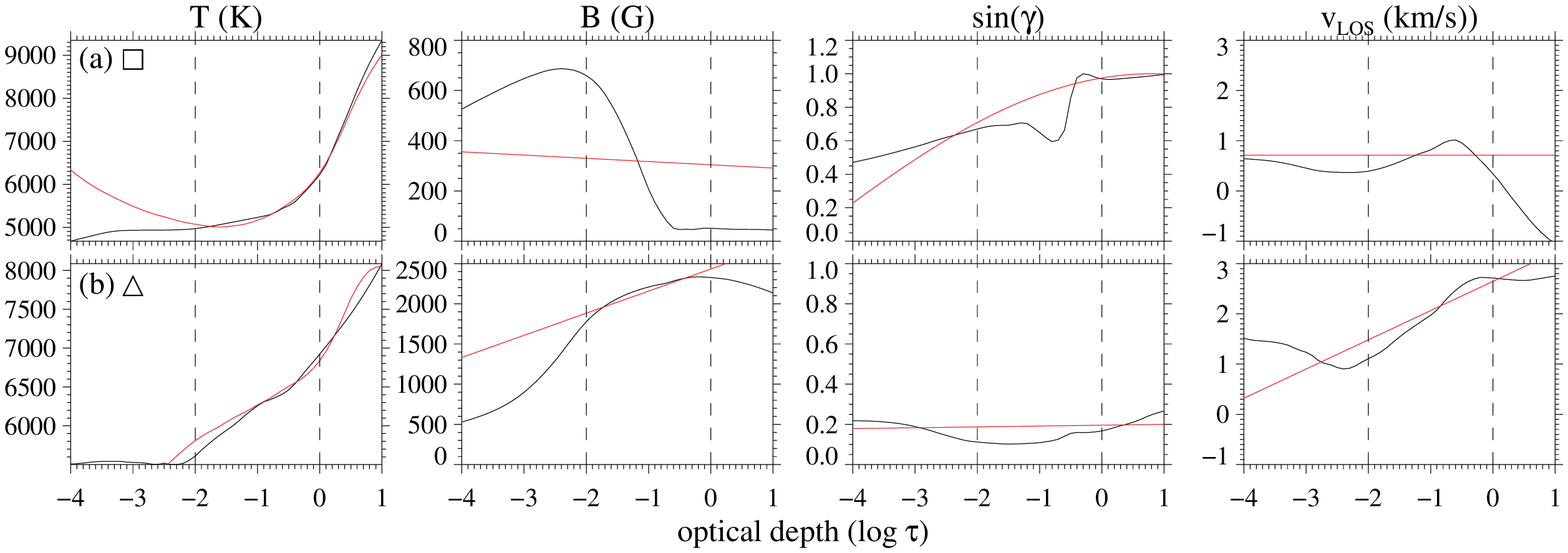}
 \centering
 \caption{Atmospheric stratification of two selected pixels extracted from area B from the simulations. Pixel (a) (\emph{top row}) corresponds to an example of a low magnetic field and is located at ($x,y$) = (0.53\arcsec, 0.13\arcsec) in Fig. \ref{Fig:sim_inv_area_b}. Pixel (b) (\emph{bottom row}) is located at ($x,y$) = (0.86\arcsec, 0.53\arcsec) and represents a pixel with a strong magnetic field. The solid black (red) line depicts the simulations (inversions). The sensitivity of the spectral lines resides mainly between the optical depths marked by the dashed vertical lines.}
 \label{Fig:atmos_tau}
\end{figure*}

In the case of map B, a case study of strong vertical fields, the inversions perform very good for the temperature, magnetic field strength, and LOS velocity, especially for $\log \tau = 0$ and $-1$. Even the field strength at $\log \tau = -2$ matches well with the simulations. As mentioned before, this is consistent with the RF seen in the lower right panel of Fig. \ref{Fig:RF}, where we still see enough sensitivity of the lines to that optical depth. The retrieved inclinations are significantly better than in the case of map A.  

A more quantitative comparison between the simulation and the inversion results is shown as scatter plots in Fig. \ref{Fig:scatterplots}. We only chose $\log \tau = -1$ because the RF is excellent for all lines at that optical depth and the match between the simulations and inversions is best. As seen in Fig. \ref{Fig:scatterplots}, the temperature and the Doppler velocity scatter plots strongly correlate 
($R > 92\%$). 
As reported before, the magnetic field is not well reproduced in the inversions ($R \sim 40\%$), when horizontal fields dominate within the FOV (area A; Fig. \ref{Fig:sim_inv_area_a}).
The most striking result to emerge from the scatter plots is the excellent correlation ($R \sim 99\%$) between the magnetic field strength when vertical magnetic fields cover most of the FOV (area B; Fig. \ref{Fig:sim_inv_area_b}). Although this might be expected, as the splitting of the line is proportional to the magnetic field, it shows the robustness and consistency of the inversions when taking into account 15 Stokes $I$ profiles. Figure \ref{Fig:scatterplots} also points out that the lower the field strength, the less accurate are the results from the inversions. This is depicted in the magnetic field strength panels, where towards the lower left corner, that is, weaker fields, the scatter is very large. 
Regarding the inferred inclinations, both scatterplots show poor matches between the simulations and the inversions. However, in area B, if we select only pixels which correspond to stronger fields, that is $B \geq 1000$\,G  (light-blue dots in Fig. \ref{Fig:scatterplots} ), we find a positive correlation. This is reflected in the blue solid line with a correlation of $R \sim 68\%$. 
On the contrary, the inferred inclinations are not trustworthy when most of the fields are perpendicularly oriented to the LOS (correlation of only $R\sim$13\,\%) or outside of stronger magnetic areas ($B < 1000$\,G).

%--------------------------------------------------------------------------
\subsection{Height variation of physical parameters}
%--------------------------------------------------------------------------

We selected two pixels from the simulation cube representing an example of strong and weak magnetic fields. Their positions are marked in Fig.~\ref{Fig:sim_inv_area_b} with a triangle (strong field) and square (weak field). 
The degraded profiles (black dots) together with their fit from the inversions (red solid line) are displayed in Fig.~\ref{Fig:profiles}. The inversion code performed excellent. The splitting of the lines, in the case of strong fields (pixel b; triangle), is clearly seen.  The inferred atmospheric stratification (red line), compared to the original simulation cube (black line), with respect to the optical depth is shown in Fig.~\ref{Fig:atmos_tau}. The response functions (Fig.~\ref{Fig:RF}) show an increased sensitivity of the spectral lines to optical depths between $\log \tau \in [0, 2]$, which are marked with vertical dashed lines in the Figure. In the case of strong magnetic fields,  Fig.~\ref{Fig:atmos_tau} lower row, the inferred stratification matches well the original one for all physical quantities except for the inclination ($\sin \gamma$), which is slightly off. In the second case (upper panels), the inversions match well the original stratification of temperature. 
The sign and order of magnitude of the velocity is similar to the original one. 
The inferred magnetic field stratification is poor, although at $\log \tau = -1$ it is accurate, which coincides with the highest sensitivity of the lines. The inferred inclination is not trustworthy. This is not surprising, since we lack the Zeeman splitting of the lines when the fields are weak. Therefore, the results of $B$ and $\gamma$ become less reliable. On the contrary, $T$ and $v$ still show good and acceptable results because they do not depend that much on the splitting of the lines.

%--------------------------------------------------------------------------
\subsection{Inversion of observations with the VTT} \label{Sect:VTTinversions} 
%--------------------------------------------------------------------------

We applied the lessons learned from the sections above to the observations. From now on we will refer only to the observed data at the VTT. 
As described in Section~\ref{Sect:observations}, a straylight profile was needed for the inversions. We performed straylight-test inversions in the quiet Sun and in the sunspot with different percentages, between 0\% and 70\%,  to find the best performance of the inversions. For large contributions (>30\%), the synthetic and observed spectra diverge. The optimal fits were achieved for a straylight between 10--20\%. Hence, we used 15\% for the inversions. We note that the straylight accounts for diffuse light, which is locally present at the telescope and in the instruments. Therefore, we cannot claim that 15\% is a good fraction for other telescopes. Furthermore, the macroturbulence was left as a free parameter, which is needed to remove the effects introduced by the unknown PSF of the telescope. 

\begin{figure}[!ht]
 \centering
 \includegraphics[width=\hsize]{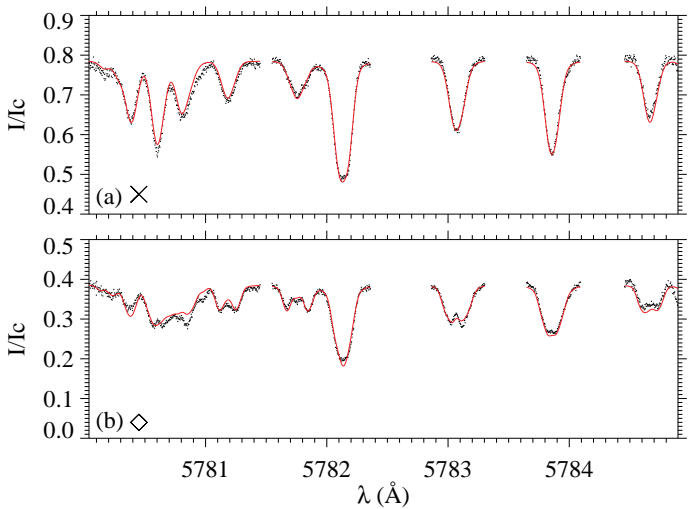}
 \caption{Performance of the inversions for two selected pixels: (a) at the penumbra and (b) in the umbra. The pixels are highlighted with a diamond and cross symbol in Fig. \ref{Fig:vtt_inversions}. The small black dots represent the observed intensity profile whereas the red solid line shows the best fit of the SIR inversions. } 
 \label{Fig:inv_profs_vtt}
\end{figure}

\begin{figure}[!h]
 \centering
 \includegraphics[width=\hsize]{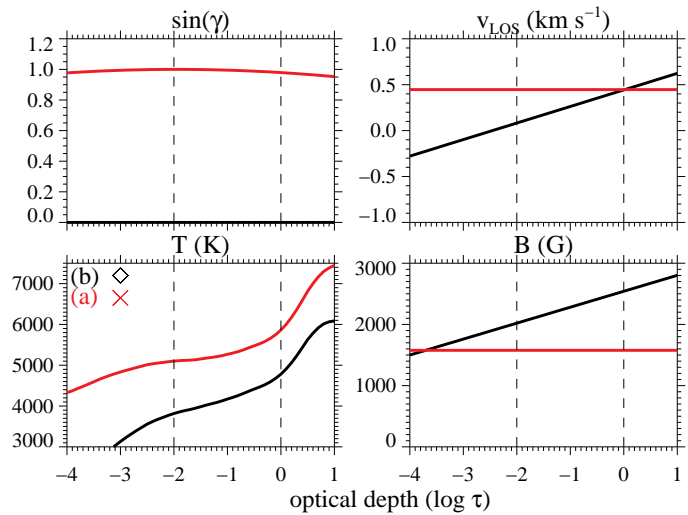}
 \caption{Atmospheric stratification of four physical parameters for two selected pixels (a) and (b) representing the penumbra (red line) and umbra (black line), respectively. Starting from the top left panel and moving clockwise: sine of the inclination, LOS velocity, magnetic field strength, and temperature. The position of the pixels is shown in Fig. \ref{Fig:vtt_inversions} with a cross and diamond symbol. The retrieved information from the inversions is most reliable inside the dashed vertical lines, that is, between $-2 \leq$\,$\log \tau$\,$\leq 0$. }
 \label{Fig:atms_inv_vtt}
\end{figure}

\begin{figure*}[!t]
\sidecaption
  \includegraphics[width=12.5cm]{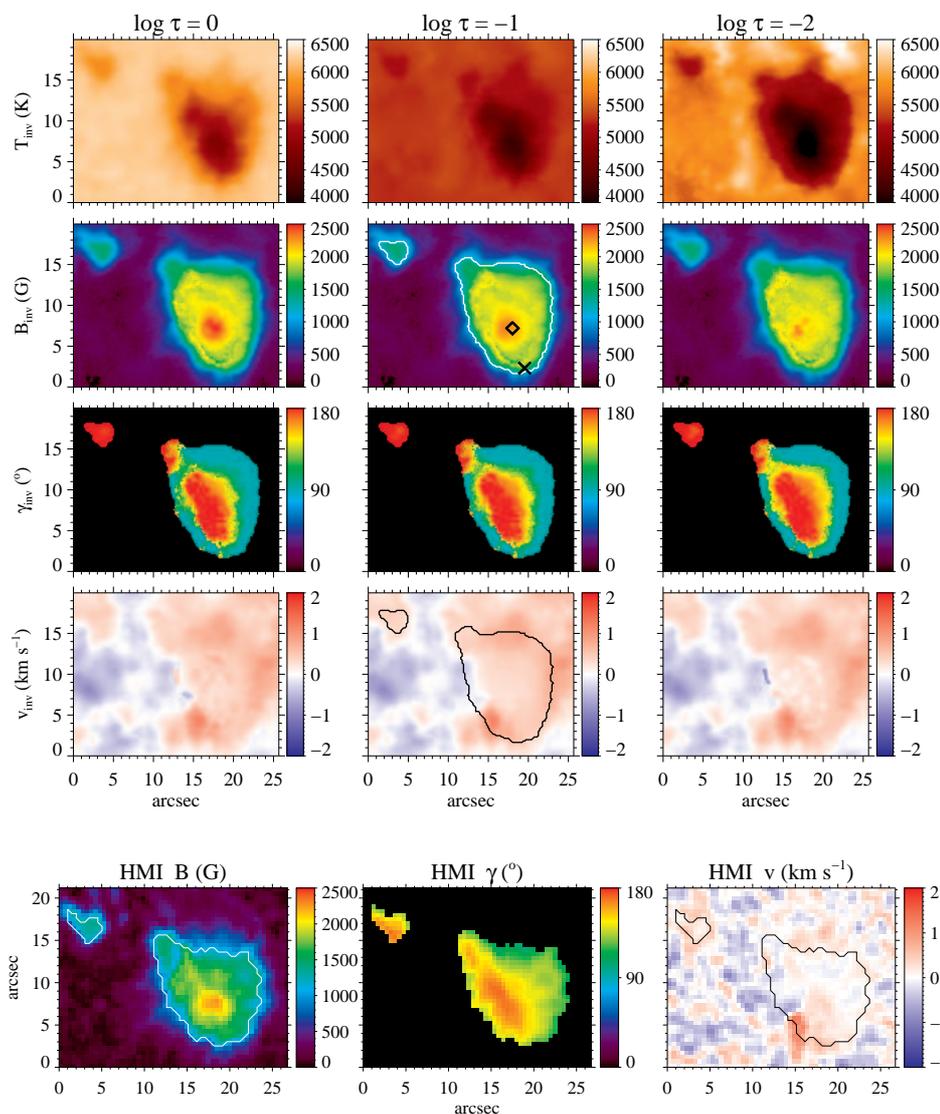}
  \caption{Results from the multi-line inversions of the VTT observations. From {\it top} to {\it bottom}: temperature $T_\mathrm{inv}$, magnetic field strength $B_\mathrm{inv}$, magnetic field inclination $\gamma_\mathrm{inv}$, and LOS velocity $v_\mathrm{inv}$. From left to right: different optical depths $\log \tau = 0, -1$, and $-2$. The white or black contours corresponds to a magnetic field of 1000\,G and was used to highlight the sunspot and pore. These contours serve as a mask for the inclination maps, showing only the inferred values for the sunspot and the pore, where the results are more reliable. The cross and diamond symbols show the location of the pixels plotted in Figs.~\ref{Fig:inv_profs_vtt} and \ref{Fig:atms_inv_vtt}. } 
  \label{Fig:vtt_inversions}
\end{figure*}

The inversions were initiated eight times, each time with a different initial-guess atmosphere as described above in the simulation inversions (Sect. \ref{Sect:InvSimulations}). 
An inspection of two individual example pixels, one at the penumbra and the other one at the umbra, is shown in Fig.~\ref{Fig:inv_profs_vtt} (a) and (b), respectively. The dots represent the observations and the solid red line the synthesized spectrum from the simultaneous inversions. The overall match is very good. Some discrepancies are mainly found in the core of the line and occasionally in the outer wings. The corresponding physical variables as a function of optical depth are depicted in Fig.~\ref{Fig:atms_inv_vtt}. The position of both pixels is highlighted with a cross and diamond symbol in the overview map in Fig.~\ref{Fig:vtt_inversions} (middle column, middle panel). As expected, it is well seen that the temperature stratification is colder in the umbra (black line in Fig.~\ref{Fig:atms_inv_vtt}) than in the penumbra (red line). Furthermore, the magnetic field is stronger and almost longitudinal in the umbra. In the penumbra we see that the field does not change with height, at least the Stokes $I$ profiles seem to indicate no changes in the range of $\log \tau \in [0,-2]$. Moreover, the inclination is close to 90$^\circ$ in the penumbra. The LOS velocities are low, in the range of 0.5 to 0.1\,km\,s$^{-1}$ between $0$\,$\geq$\,$\log \tau$\,$\geq$\,$-2$, for both pixels.

The results, for three fixed optical depths ($\log \tau \in [0,-2]$), appear in Fig.~\ref{Fig:vtt_inversions}. 
We manually disambiguated pixel by pixel the polarity of the magnetic field lines by using as a reference the inclination map of HMI (Fig. \ref{Fig:HMIinc}). Overall, our inversions exhibit a remarkable performance at highly magnetized atmospheres, as proven by the examination of the numerical simulations (Fig.~\ref{Fig:scatterplots}). A more detailed analysis of the results will be discussed in Sect.~\ref{Sect:discussion}. 

%---- Figure -----------------------------------------------------------------
%-------------------------------------------------------------------------------

%--------------------------------------------------------------------------
\subsection{Magnetic field and inclination inferred from the Helioseismic and Magnetic Imager}
\begin{figure*}[!t]
 \centering
\sidecaption
  \includegraphics[width=12.5cm]{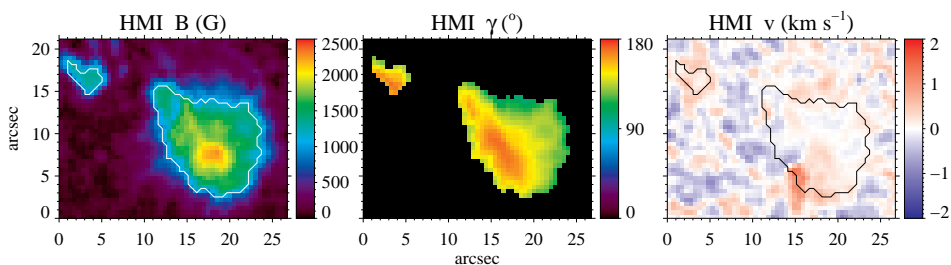}
 \caption{HMI Milne-Eddington inversions for a region-of-interest on 2018 September 30 at 10:10\,UT. The white or black contour outlines a mask with pixels which fulfill B$\geq$1000\,G. The LOS  inclination (middle panel) only shows pixels within the aforementioned mask. }
 \label{Fig:HMIinc}
\end{figure*}

%--------------------------------------------------------------------------
The HMI full-disk vector-field data are represented in Fig.~\ref{Fig:HMIinc}. 
The sunspot shows a magnetic field strength in the umbra of up to 2500\,G, whereas the outer-surrounding penumbra decreases until about 1000\,G. The LOS inclination panel shows longitudinal fields in the umbra and, towards the penumbra, the field gradually gets more transverse. HMI provides the correct polarity, that is, orientation of the field lines with respect to the line-of-sight, which is not present in our VTT observations which lack polarimetric measurements. Therefore, there is an ambiguity in our inferred inclination, which arises from the fact that Stokes $I$ is not sensitive to the polarity of the field (see also Sect.\ref{Sect:methods}). HMI magnetograms solve this ambiguity. 

%--------------------------------------------------------------------------
\section{Discussion and summary} \label{Sect:discussion}
%--------------------------------------------------------------------------

We explored the potential of the spectral range in the vicinity of the central \ion{Cr}{I} line at 5781.75\,\AA\ to infer physical parameters, such as temperature, magnetic field, and Doppler velocities, without polarimetry. The chromium lines were already used to discover the magnetic field in sunspots more than a century ago by \citet{hale06}. The \ion{Cr}{i} line at 5783\,\AA\ was used in the past to infer the magnetic field vector \citep{balthasar09}. However, this spectral window offers even a higher diagnostic potential than in those previous studies. It contains 15 suitable spectral lines which arise from the photosphere of the Sun. All of the lines were sensitive to the magnetic field. \citet{balthasar12} showed that inverting simultaneously 15 lines, including polarimetry, yields good results with the SIR code. Here we test such an inversion strategy in an unexplored spectral range and only using intensity spectra. 

We determined first the oscillator strengths ($\log gf$) of all lines (Sect. \ref{Sect:loggf}). Furthermore, with the use of the inversion code SIR, we examined for which optical depths the spectral lines deliver information (using response functions, see Sect. \ref{Sect:RFs}) and what physical parameters are trustworthy. To that end, we synthesized the spectral lines arising from a well-known simulation cube \cite[snapshot 385,][]{Carlsson2016}. 
The Stokes-$I$ synthesized spectra were then degraded to obtain more realistic conditions and inverted with SIR, to derive the best inversion strategy for SIR. In addition, the reliability of only Stokes-$I$ inversions to retrieve the correct physical parameters of the atmosphere was checked. 
The information extracted from the spectra yields a good approximation of the original temperature, magnetic field, and LOS velocity stratification from the simulations. We noticed a substantially better performance of the inversions where strong and vertical magnetic fields ($B$>1000\,G) were dominating (see scatter plots in Fig.~\ref{Fig:scatterplots}). This can be naturally explained by the Zeeman splitting pattern which becomes clearly visible with strong magnetic fields and improves the inversion performance. In addition, the height stratification of the magnetic field also plays an important role, which in our case was monotonically increasing or decreasing with optical depth in the area of strong vertical fields. In contrast, when a horizontal and non-monotonic height-stratification of the magnetic field is present, the inversions are less efficient and splitting of the lines is absent. 

Overall, the results showed that we can reliably infer the atmospheric stratification of the temperature and LOS velocity using the 15 intensity profiles simultaneously. Furthermore, the magnetic field strength and its inclination can be retrieved, however, with some limitations depending on the strength of the magnetic field ($B$>1000\,G). Keeping this in mind, we inverted observations of the VTT including a small sunspot. 
The results will be briefly discussed in the next section.

%--------------------------------------------------------------------------
\subsection{Physical interpretation of the VTT inversions}
%--------------------------------------------------------------------------

In the current manuscript, we pursued a challenging approach and inverted only the intensity of a large set of photospheric spectral lines to retrieve not only the thermodynamics of an active region but also its magnetic structure.

For the application of the Stokes $I$ only inversion technique, we focus on the examination of the inversions at an optical depth where the results are most reliable ($\log\tau$=$-1$, see Sect. \ref{Sect:VTTinversions}). Figure \ref{Fig:vtt_inversions} shows results of the inversion of the observed active region NOAA 12723. They are consistent with our current knowledge about the properties of active regions. The darkest regions of the sunspot umbra correspond to highly-magnetized atmospheres. There, the temperature exhibits the lowest values (4000\,K), which are associated with strong (around 2500\,G) and mostly longitudinal magnetic fields. At increasing radial distances from the center of the umbra, the temperature increases, and the magnetic field strength is reduced. The inclination of the magnetic field also increases with the radial distance, reaching approximately transverse magnetic fields at the outer boundary of the region where the field inclination is plotted (where the magnetic field strength is 1000\,G).
Indications of the Evershed flow are found around the sunspot (see lower panels of Fig.~\ref{Fig:vtt_inversions}). In summary, our inversions reveal the well-known correlation between the intensity of the sunspot atmosphere and its temperature and inclination and the anti-correlation of these quantities with the magnetic field strength.

%--------------------------------------------------------------------------
\subsection{Comparison between VTT and HMI}
%--------------------------------------------------------------------------

A general comparison between the VTT and HMI results is presented below.
The HMI results are based on Milne-Eddington inversions \citep{borrero2011}. While the HMI instrument observes the four Stokes parameters, the VTT acquires only intensity spectra with high spectral resolution. When comparing the magnetic field strength panels of the VTT inversions (Fig.~\ref{Fig:vtt_inversions}) with the HMI inversions (Fig.~\ref{Fig:HMIinc}) we find a similar overall tendency. 
The results agree well for field strengths along the LOS above 1000\,G, which is in line with what was found in the simulations.
The best match is visually found for an optical depth of $\log \tau = -1$ and we concentrated on this optical depth for the HMI comparison. The central umbra (red area) reaches similar field strengths of up to 2500\,G. However, surrounding the strong umbra, the HMI field strength drops to values between 1500--2000\,G whereas in the FOV of the VTT the fields are predominantly stronger, around 2000\,G. The magnetic field inclination from VTT overall resembles the one from HMI, showing in the middle umbra longitudinal fields and towards the borders of the sunspot, at the penumbra, transverse fields. The LOS velocities for the VTT inversions are much smoother than the results from HMI, which show the granular pattern.  

The small pore in the upper left corner shows field strengths of the same order of magnitude in both instruments, HMI and VTT, between 1000\,G and 2000\,G. Regarding the inclination, VTT only exhibits fields which are aligned along the LOS. HMI shows a smooth gradient from longitudinal to more transverse fields. One possible explanation is that the inclination depends strongly on the central $\pi$-components. Weak components are not recognized within the noise, thus the SIR code determines inclination values close to 0$^\circ$ or 180$^\circ$. When the component is strong, $\sin \gamma$ varies only slightly, and the result is close to 90$^\circ$, and again, noise plays a role.
The LOS velocities overall coincide in both instruments, showing slight redshifts in the pore. 

The comparison between the HMI and VTT inversions gives further confidence in our method and the inversions.A more detailed comparison between both instruments is not possible since there are too many factors, such as different types of instrument, times, wavelengths, seeing, methodology, etc, affecting a reliable one-to-one comparison.

%--------------------------------------------------------------------------
\subsection{Prospects of the method}
%--------------------------------------------------------------------------
We have proven that the spectral range around the central \ion{Cr}{I} line at 5781.75\,\AA\ is well suited to study the magnetic field in the lower photosphere. In addition, physical parameters such as temperature and Doppler velocities across different optical depths can be inferred up to $\log \tau \sim -2$. The 15 lines in this spectral region provide information from similar atmospheric heights. The advantage of combining all of them in a single multi-line inversion was shown in this study. On the one hand, the inversions provided smooth solutions and homogeneous distributions of the physical parameters. No smoothing or restart of the inversions with cleared input atmospheres was needed. This makes this inversion strategy efficient and relatively easy to implement into existing data-reduction and \mbox{-analyses} pipelines such as sTools \citep{stools}. On the other hand, no polarimetry was required to retrieve the magnetic field and its inclination in strongly magnetized areas. This greatly simplifies the observations, not only because of lower demands on the instrumentation, but also because of the absence of the need for an elaborated polarimetric calibration. In addition, concentrating only on intensity profiles has the advantage of fast cadence maps which are hard to achieve with polarimetry since state-of-the-art integral field units are still limited to small FOVs. Such fast cadence maps are necessary for fast-evolving phenomena on the Sun or for wave analyses. Nevertheless, we emphasize that inversions including polarimetry are surely more accurate and less limited to extract the magnetic field vector. The approach shown here is however an excellent method to obtain this information when no polarimetry is available. In addition, we provide here all the necessary ingredients to exploit the spectral lines between 5780.1 and 5784.6\,\AA. Future infrastructures such as the European Solar Telescope \citep{EST} or telescopes with a spectrograph, which lack polarimetry, will benefit from the present study. 

Motivated by the recent scientific results derived from observations with the VTT Echelle spectrograph \citep{verma20,kontogiannis20,abbasvand20}, AIP devised an upgrade of the facility cameras available for spectroscopic observations. The Fast Multi-Line Universal Spectrograph (FaMuLUS) integrates four large-format (8k $\times$ 6k) CMOS cameras, which scan a FOV of $240\arcsec \times 120\arcsec$ in about one minute. The effective spatial resolution (image scale of 0.36\arcsec\ pixel$^{-1}$) is well matched to SDO HMI and AIA data. The operation mode for on-disk observations includes the chromospheric H$\alpha$ and H$\beta$ lines, information of the photospheric magnetic field is obtained in the \ion{Cr}{i} 5782~\AA\ range, and the non-Zeeman ($g=0$) \ion{Fe}{i} at 7090~\AA\ line provides reliable information of the photospheric LOS velocity. The SIR inversions described in this study are accompanied by Cloud Model inversions for the H$\alpha$ and H$\beta$ lines \citep[for example, with VTT data,][]{kuckein16, Dineva2020}. Other operation modes for limb observations of prominences are also possible. Thus, FaMuLUS will deliver a comprehensive set of atmospheric properties covering a broad variety of science cases. Commissioning of the FaMuLUS camera system is scheduled for the second half of 2021.

%===============================================================================
%    Acknowledgements
%===============================================================================

\begin{acknowledgements}
We would like to thank the anonymous referee for his/her comprehensive review and valuable suggestions.
The Vacuum Tower Telescope (VTT) at the Spanish Observatorio del Teide of the
Instituto de Astrof{\'i}sica de Canarias is operated by the German consortium of the Leibniz-Institut fur Sonnenphysik (KIS) in Freiburg, the Leibniz-Institut fur Astrophysik Potsdam (AIP), and the Max-Planck-Institut fur Sonnensystemforschung (MPS) in G\"ottingen. 
Funding from the Horizon 2020 projects SOLARNET (No 824135) and ESCAPE (No 824064)
is gratefully acknowledged. 
This work was supported by grant DE 787/5-1 of the Deutsche Forschungsgemeinschaft (DFG). CQN was supported by the EST Project Office, funded by the Canary Islands Government (file SD 17/01) under a direct grant awarded to the IAC on ground of public interest, and this activity has also received funding from the European Union’s Horizon 2020 research and innovation programme under grant agreement No 739500. TF acknowledges financial support from the State Research Agency (AEI) of the Spanish Ministry of Science, Innovation and Universities (MCIU) and the European Regional Development Fund (FEDER) under grant with reference PGC2018-097611-A-I00. MV was supported by grant VE~1112/1-1 of the DFG. This research has made use of NASA's Astrophysics Data System.

\end{acknowledgements}

%===============================================================================
%    BIBLIOGRAPHY
%===============================================================================

\bibliographystyle{aa}
\bibliography{aa-jour,biblio}

\end{document}